\documentclass[a4,12pt]{article}

\usepackage[T1]{fontenc}
\usepackage{times}

\usepackage{graphicx}
\usepackage{color}

\usepackage{subfigure}

\usepackage[final]{prelim2e}

\usepackage{amsmath}

\usepackage{textcomp}

\newcommand{\mm}{\textsc{Mathematica$^\textrm{\textregistered}$ }}
\newcommand{\ml}{\textsc{Matlab$^\textrm{\textregistered}$ }}

\title{Comparison of two different implementations of a finite-difference-method for first-order pde in \mm and \ml}

\author{\textbf{Heiko Herrmann}\footnotemark[1]\ \  \footnotemark[2] 
\and Gunnar R\"uckner\footnotemark[1]}

\date{}

\begin{document}
\maketitle
\setlength{\parindent}{0em}
\begin{abstract}
\noindent 
In this article two implementations of a symmetric finite difference algorithm (ftcs-method) for a first-order partial differential equation are discussed. The considered partial differential equation discribes the time evolution of the crack length ditribution of microcracks in brittle materia.
\end{abstract}
\renewcommand{\thefootnote}{\fnsymbol{footnote}}
\footnotetext[1]{Technische Universit\"at Berlin -- Institut f\"ur Theoretische Physik -- Sekr.:  PN 7-1 -- Hardenbergstr. 36 -- 10623 Berlin -- Germany}
\footnotetext[2]{Correspondend autor: Heiko Herrmann}
\renewcommand{\thefootnote}{\arabic{footnote}}

\section{Physics and analytical solution}
The growth rate of microcracks in a brittle material can be discribed by a mesoscopic equation. Here the specialized version for uniaxial loading is presented.
\begin{eqnarray}
\frac{\partial f(l,t)}{\partial t} & = & -\frac{1}{l^2} \frac{\partial l^2 v_l(l,t) f(l,t)}{\partial l},
\label{eq_mored}
\end{eqnarray}
$f(l,t)$ is the distribution function for the crack length $l$ at time $t$, $v_l = \dot l$ is the growth velocity of the cracks. A Rice-Griffith-like dynamic is assumed for crack growth, which gives 
\begin{eqnarray}
\dot l & = & \begin{cases} - \alpha' + \beta' l \sigma(t)^2 & \textrm{, if } \alpha' \leq \beta' l \sigma^2 \\ 0 & \textrm{, otherwise} \end{cases}
\end{eqnarray}
The theory is given in detail in \cite{VanPa02}. For an exponential (or a step-wise) initial condition and constant loading speed it is possible to give an exact analytical solution, which is also presented in \cite{VanPa02} and looks like:
\begin{eqnarray}
f(l,t) & = &
\begin{cases} l^{-2} e^{-\frac{\beta' v_\sigma^2}{3}t^3} F\left(l e^{-\frac{\beta' v_\sigma^2}{3}t^3} + \frac{\alpha'}{(9 \beta' v_\sigma^2)^{1/3}} \Gamma(1/3,0, \frac{\beta' v_\sigma^2}{3}t^3) \right) & \textrm{, if}\quad \alpha' \leq \beta'v_\sigma^2 l t^2 \\
& \\
f(l,0) & \textrm{, otherwise}.
\end{cases}
\end{eqnarray}

\begin{figure}[htbp]
\subfigure[step-wise initial condition]{
\includegraphics[width=6cm]{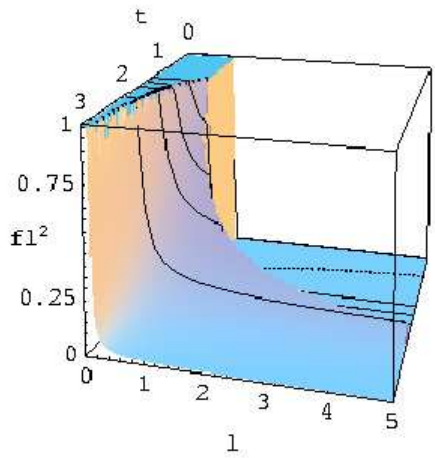}
}
\subfigure[step-wise initial condition]{
\includegraphics[width=6cm]{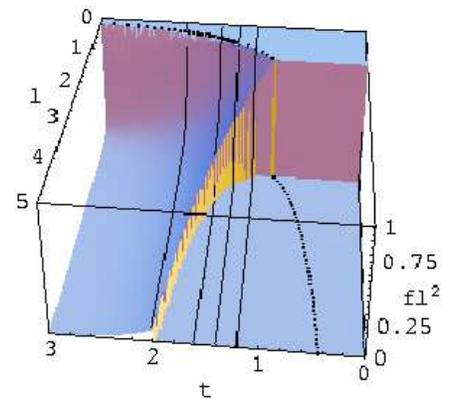}
}
\subfigure[exponential initial condition]{
\includegraphics[width=6cm]{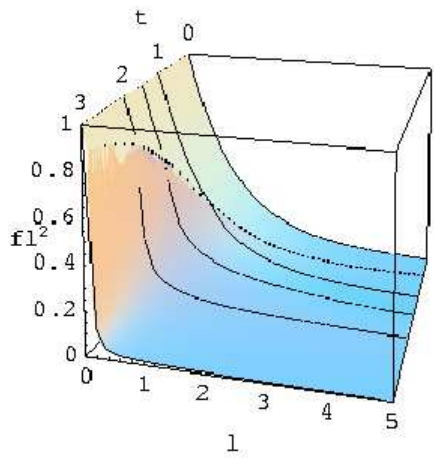}  
}
  \caption{Analytical solution for crack lenght distributuion. Shown is $f(l,t) l^2$ over $l$ and $t$.}
  \label{fig:anal}
\end{figure}
\clearpage

\section{The numerical algorithm}
The partial differential equations are first order in time and crack length. Two different algorithms, upwind and fcts (\textbf{f}orward \textbf{t}ime \textbf{c}entered \textbf{s}pace), have been tested. The symmetric algorithm (ftcs-method) is in this case a bit more stable than the upwind, which is somewhat astonishing. Both algorithms can be found in \cite{NRC}.\\
In the symmetric algorithm $f(l,t+1)$ is calculated from $f(l-1,t)$, $f(l,t)$ and $f(l+1,t)$ by
\begin{eqnarray}
  f(l,t+1) & = & f(l,t) \left(1-\left(3 \beta \sigma^2(t) - \frac{2 \alpha}{dl\ l}) dt\right)-\left(dl\ l \beta \sigma^2(t)\right) - \alpha\right) \times \nonumber \\
&& \times \frac{dt}{2 dl} \left(f(l+1,t)-f(l-1,t)\right)
\end{eqnarray}
This is what the main part of the implementation in \mm looks like:
%% cell bracket markieren, dann cell menue -> convert to -> InputForm
%% danach geht kopieren mit markieren und paste in verbatim-Umgebung
{\small
\begin{verbatim}
For[t = 1, t < TMAX, t++, For[l = 1, l < LMAX, l++, 
   If[l*dl*(t*dt)^2*be - al > 0, 
    If[l == 0, f[l, t + 1] = f[l, t], 
     f[l, t + 1] = 
          f[l, t]*(1 - (3*be*(t*dt)^2 - (2*al)/(dl*l))*dt) - 
          (dl*l*be*(t*dt)^2 - al)*dt/dl*(f[l + 1, t] - f[l - 1, t])], 
    f[l, t + 1] = f[l, t]
    ]
   ]
]
\end{verbatim}
}
This is what the main part of the implementation in \ml looks like:
{\small
\begin{verbatim}
for t=1:1:TMAX-1
  for l=1:1:LMAX-1
    if (l.*dl.*sigma(t).*beta-alpha > 0)
      f(l,t+1)=f(l,t).*(1-(3.*beta.*(sigma(t)).^2 - ...
(2.*alpha)./(dl.*l)).*dt)-(dl.*l.*beta.*(sigma(t)).^2 - ...
alpha).*dt./(2*dl).*(f(l+1,t)-f(l-1,t));
    else
    f(l,t+1)=f(l,t);
    end
  end
end
\end{verbatim}
}

\newpage
\section{Results obtained by \mm}
\mm is a general purpose computer algebra system (cas) by Wolfram Research Inc..
\begin{figure}[htbp]
\centering
\subfigure[step-wise initial condition]{
\includegraphics[width=6cm]{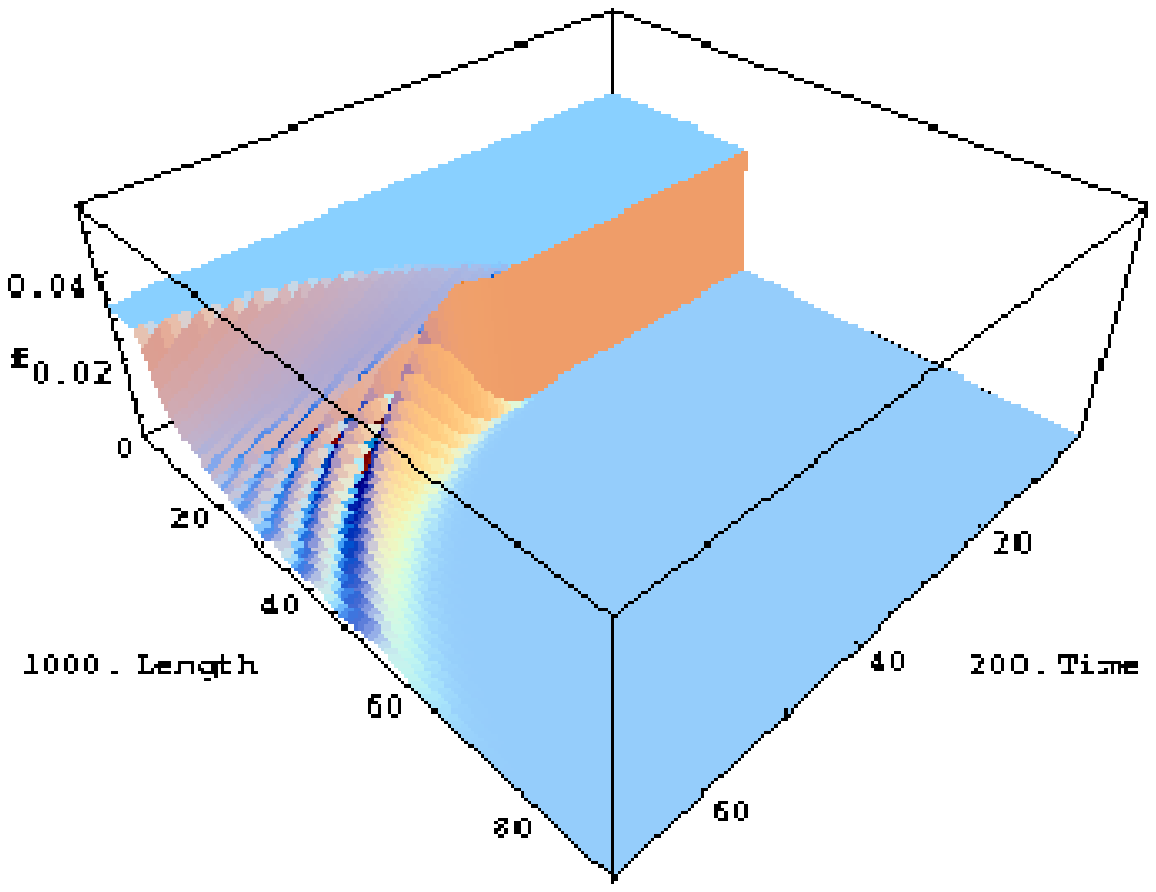}
}
\subfigure[exponential initial condition]{
\includegraphics[width=6cm]{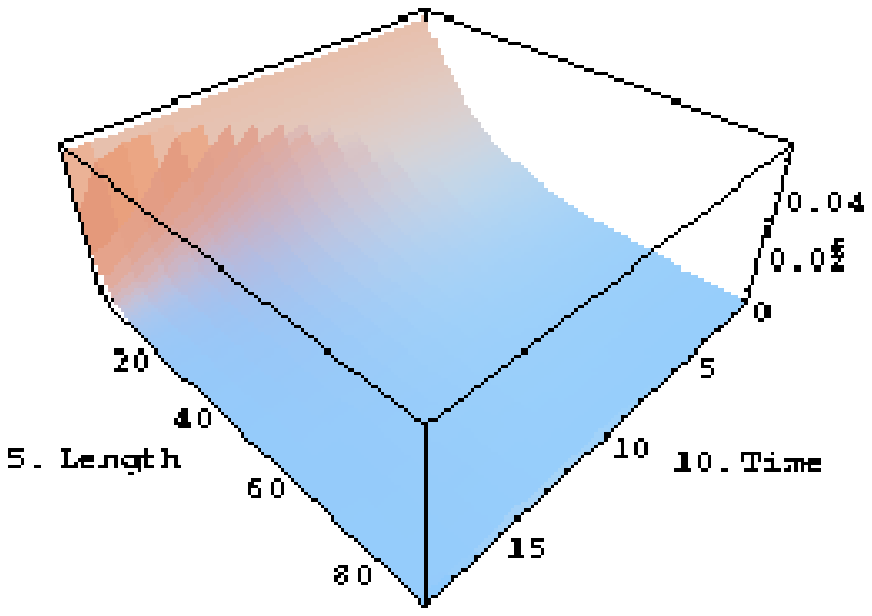}
}
  \caption{Numerical solution for crack lenght distributuion calculated with \mm}
  \label{fig:math}
\end{figure}
As one can see the solution shows some wave-like efects, which can be interpreted as a sign for numerical instability. This instability is a result of the discontinous initial condition.
\clearpage

\section{Results obtained by \ml}
\ml is a tool for numerical mathematics, especially designed for matrix manipulation, by The MathWorks Inc..
\begin{figure}[htbp]
\centering
\subfigure[step-wise initial condition]{
\includegraphics[width=6cm]{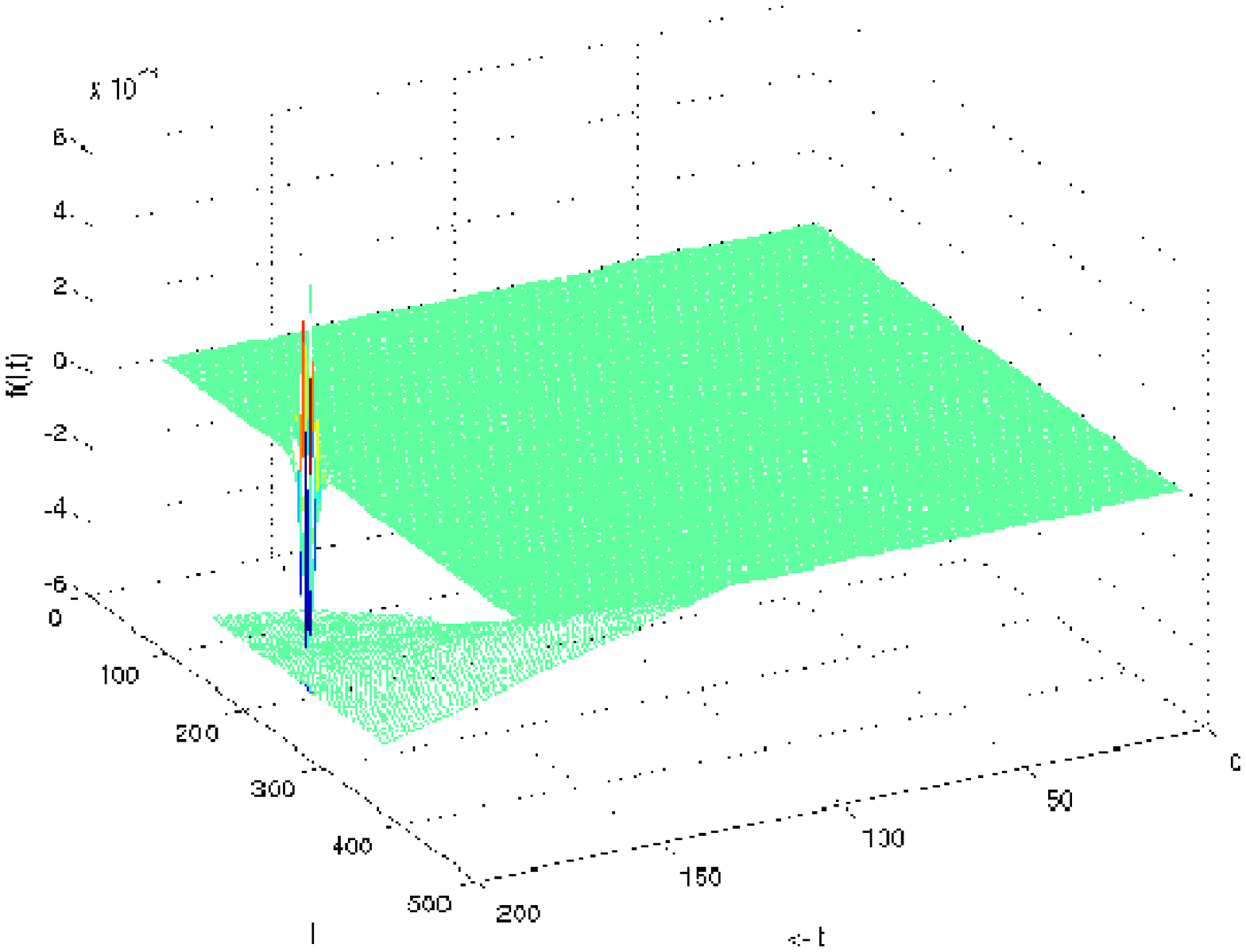}
}
\subfigure[exponential initial condition]{
\includegraphics[width=6cm]{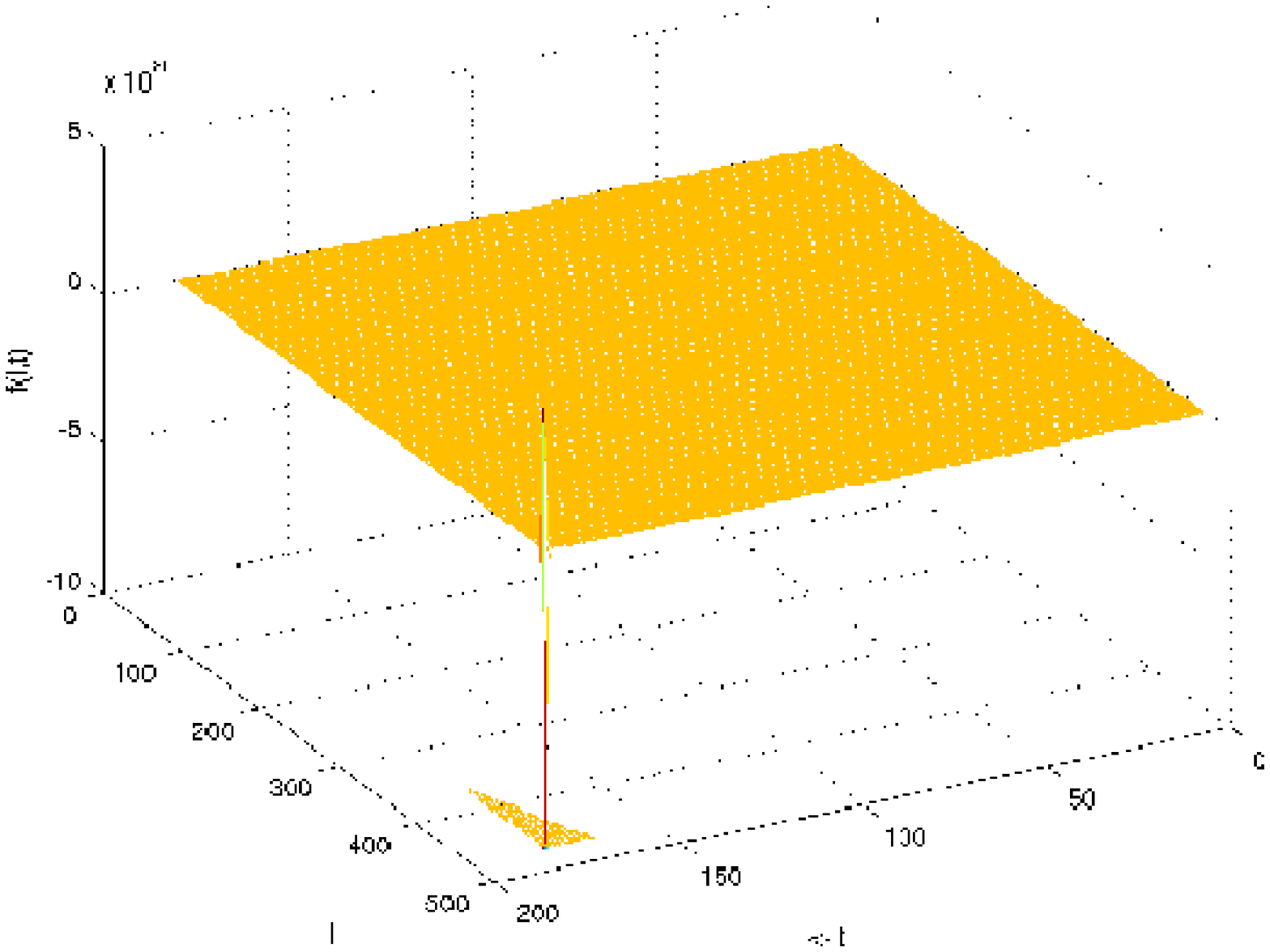}
}
  \caption{Numerical solution for crack lenght distributuion calculated with \ml}
  \label{fig:matl}
\end{figure}
Here in both cases huge instabilities occur. As soon as some cracks start growing the numerical error goes to infinity. The following pictures show an extract of the above pictures.
\begin{figure}[htbp]
\centering
\subfigure[step-wise initial condition]{
\includegraphics[width=6cm]{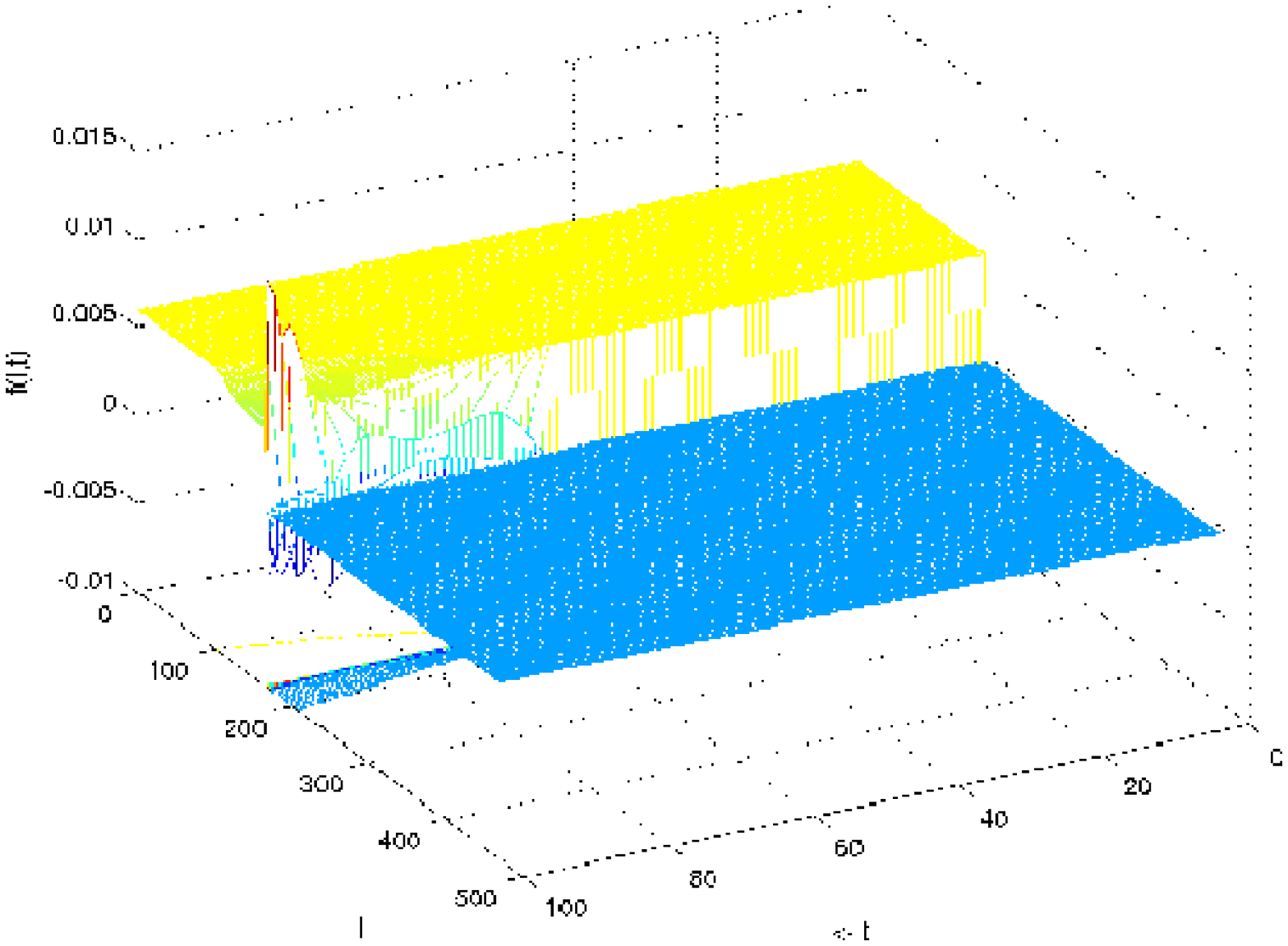}
}
\subfigure[exponential initial condition]{
\includegraphics[width=6cm]{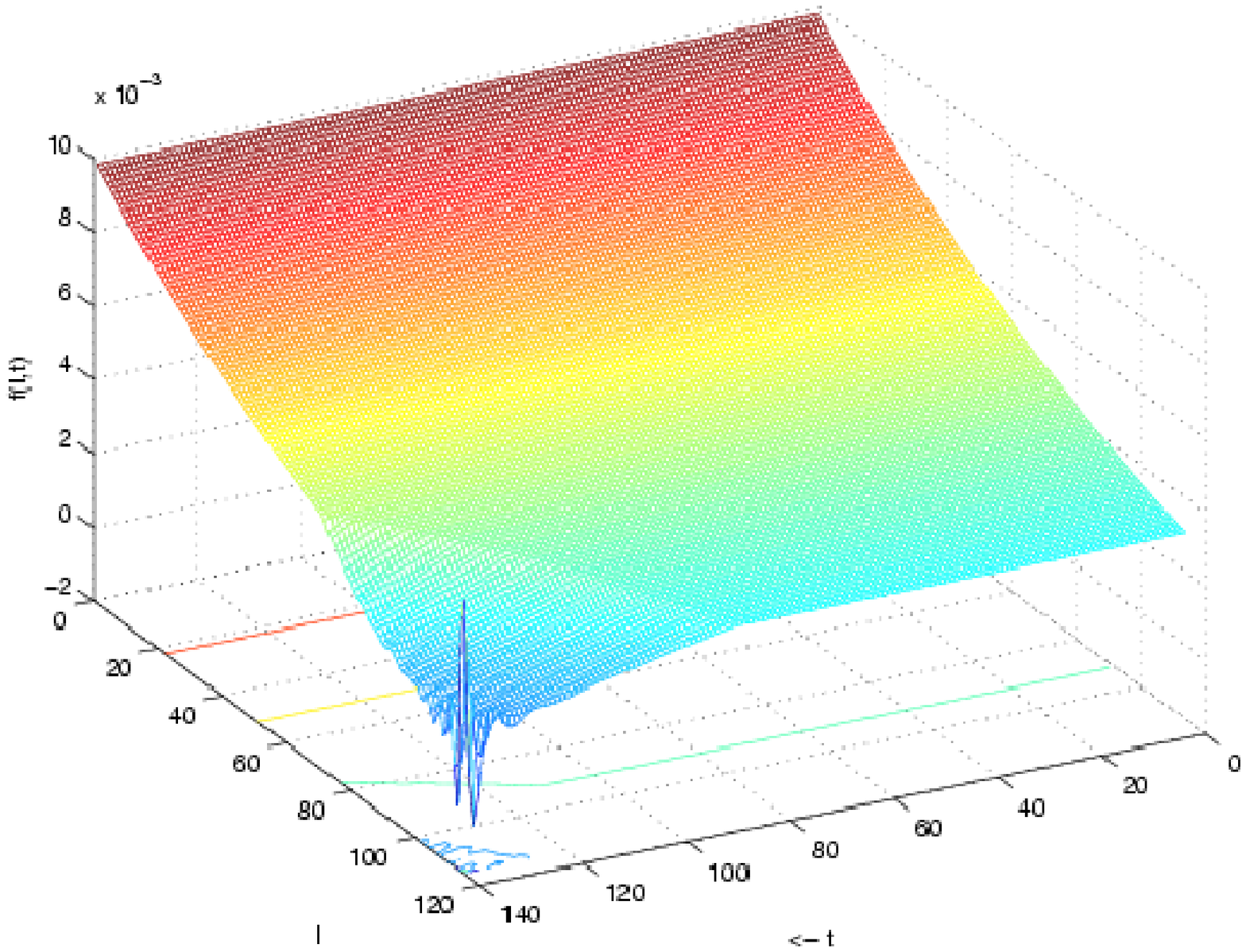}
}
  \caption{Numerical solution for crack lenght distributuion calculated with \ml (extract of figure \ref{fig:matl})}
  \label{fig:mat2}
\end{figure}
\clearpage

\section{Comparison of the results and conclusion}
Obviously the results, obtained with both programs, show huge errors. But in contrast to the results calculated with \ml, one can use the results obtained with \mm at least for some rough predictions.\\
The difference in the two software packages, used for these simulations, is that \ml uses floating-point variables of precision ``double'' (16 byte), whereas \mm is capable of both numerical and symolic computation. Therefore it is possible that \mm uses a much higher precision to perform some of the operations than \ml. %%Unfortunately, as with most commercial programs, it is not possible to determine what \mm is doing \emph{exactly}.\\

\section*{Acknowledgement}
We thank Dr. Christina Papenfu{\ss} and Dr. Peter V\'an for discussions. Financial support by the DAAD, OTKA and by the VISHAY Company, 95100 Selb, Germany, is greatfully acknowledged.

\bibliographystyle{unsrt}
\bibliography{numrg}
\end{document}